# Programmable collective behavior in dynamically self-assembled mobile microrobotic swarms

Berk Yigit[#], Yunus Alapan[#], and Metin Sitti[*]

Physical Intelligence Department, Max Planck Institute for Intelligent Systems, 70569 Stuttgart, Germany

[*] E-mail: sitti@is.mpg.de

[#] The authors contributed equally and share the first authorship.

**Abstract**

Collective control of mobile microrobotic swarms is indispensable for their potential high-impact applications in targeted drug delivery, medical diagnostics, parallel micromanipulation, and environmental sensing and remediation. Lack of on-board computational and sensing capabilities in current microrobotic systems necessitates use of physical interactions among individual microrobots for local physical communication and cooperation. Here, we show that mobile microrobotic swarms with well-defined collective behavior can be designed by engineering magnetic interactions among individual units. Microrobots, consisting of a linear chain of self-assembled magnetic microparticles, locomote on surfaces in response to a precessing magnetic field. Control over the direction of precessing magnetic field allows engineering attractive and repulsive interactions among microrobots and, thus, collective order with well-defined spatial organization and parallel operation over macroscale distances (~1 cm). These microrobotic swarms can be guided through confined spaces, while preserving microrobot morphology and function. These swarms can further achieve directional transport of large cargoes on surfaces and small cargoes in bulk fluids. Described design approach, exploiting physical interactions among individual robots, enables facile and rapid formation of self-organized and reconfigurable microrobotic swarms with programmable collective order.

**Keywords:** Swarm, microrobotics, programmable self-assembly, collective systems, reconfigurable systems

Operation of functional microrobots as individual units, as well as in swarms, has the potential to revolutionize manipulation of the microscopic world. Various microrobots, actuated via magnetic, electrical, and biological means, have been developed and proved significant potential in cargo transportation, micromanipulation, microsurgery, sensing, environmental remediation, and drug delivery applications at the microscale (< 100 μm)[1-8]. However, considering the size difference among microrobots and their targets, such as tissues and organs, capacity of an individual microrobot would be insufficient for achieving desired effects on the macroscale (~1 cm). Therefore, many real-world medical applications would require microrobots to operate in parallel to amplify functional throughput, without hindering the motility and function of one another[9]. Furthermore, microrobotic collectives would be able to cooperate in large groups to generate higher order functionalities that is required to accomplish complex tasks, going well beyond the capability of an individual microrobotic unit[9]. In order to achieve this goal, physical communication among microrobots through local physical interactions are required in the absence of on-board computational and sensing capabilities[10-13]. Therefore, programming collective interactions at the microscale requires fundamental understanding and engineering of physical interactions among microrobots, which remains to be a significant scientific challenge.

Swarming synthetic and bio-hybrid microrobots have been investigated for micromanipulation and medical applications[7,14-20]. In general, the main strategy for controlling swarms relies on the motile response of microrobotic units to remotely controlled global fields, such as magnetic fields. Magnetotactic bacteria swarms were used to manipulate microobjects to assemble larger structures, although swarm dynamics was governed by global magnetic field gradients instead of interactions among bacteria[16]. Similarly, synthetic helical microswimmer aggregates, actuated with global magnetic fields, were shown to enhance real-time signals for medical imaging modalities[7,17]. Recently, swarms of reconfigurable microrobots formed using magnetic micro/nanoparticles were used as hyperthermia agents and for mechanical lysis of fibrin gels[18,19]. Despite all these advances in fabrication of microrobot swarms, physical interactions among microrobots are still yet to be elucidated for engineering collective order, including spatial organization of microrobots and their parallel operation.

Recent advances in bottom-up magnetic colloidal assembly have enabled engineering microrobots with controlled size, shape, and function[21-24]. However, collective operation of such magnetically self-assembled microrobots in large numbers suffer from uncontrolled aggregation of robots due to magnetic attractions among individuals. Here, we show that microrobotic swarms with well-defined collective order can be fabricated via dynamic self-assembly of magnetic particles into mobile microrobotic linear chains that communicate with each other through engineered magnetic interactions. Such self-assembled microrobots are generated simultaneously in large numbers and their propulsion near a solid surface can be driven and controlled using precessing global magnetic fields. Microrobot swarms can be reversibly assembled and disassembled by applying and removing the external magnetic field on-demand. Once assembled, precise control over magnetic field allows engineering the collective behavior of microrobots through tuned pairwise magnetic dipole-dipole interactions. We show that these microrobotic swarms can be guided through confined environments, without compromising their structural and functional integrity. Furthermore, these motile microrobotic arrays can be used to generate directional flows and transport cargoes of various sizes (1-20 μm) on a surface and inside a bulk fluid. This system demonstrates a blueprint towards the next generation of rapid, reconfigurable, and reversible assembly of functional microrobotic swarms with well-defined collective order.

## Results

**Dynamic self-assembly of mobile microrobotic linear chains and their pairwise interactions**

Microrobots were constructed using superparamagnetic microparticles (around 5 µm diameter) and applied magnetic fields, resulting in self-assembly of microparticles into linear chain structures **(Figs. 1a, b)**. Induced magnetic dipole moments of superparamagnetic microparticles follow the orientation of the applied global magnetic field (***B***) and particles interact with each other via their dipoles. Therefore, when a rotating magnetic field following a conical path with a semi-cone angle ($\Psi = 70°$) is applied at a small angular frequency ($\omega/2\pi = 1$ Hz), particles attract each other along their magnetic dipoles and assemble into chains **(see SI for details, Movie S1)**. Once assembled, chains revolve about their centers following the conical path of the magnetic field, which can be adjusted by varying $\Psi$ ($\Psi = 0°\text{-}90°$) **(Fig. 1b)**. Upon application of a tilt angle ($\vartheta$) to the precession axis of the magnetic field, self-assembled microrobots locomote on the surface, while revolving around their precession axis **(Fig. 1c, Movie S1)**. Upon removal of the applied magnetic field, chains disassemble into individual beads **(Movie S1)**.

Time-dependent forces arise among microrobot chains due to magnetic dipole-dipole interactions. The magnitude and direction of magnetic forces are determined by the precession and tilt angles of the precessing magnetic field, which results in a net attractive or repulsive interaction among chains when averaged over a rotational cycle **(Fig. S1)**. We performed our experiments at various combinations of tilt and precession angles that represent the broad workspace of field configurations $0° \leq \Psi, \vartheta \leq 90°$ **(Fig. S1)**, while focusing on the cases where the sum of tilt and precession angles does not exceed 90° to prevent chain fragmentations due to direct collisions among chains and surface **(Figs. 1d, e)**. Numerical analyses of normalized magnetic dipolar interaction forces between two chains for different tilt and precession angle combinations showed almost homogeneous repulsive force vectors for all radial directions (β = 0°-360°) for configurations of Ψ = 45°-ϑ = 22° and Ψ = 30°-ϑ = 30°. For configurations of Ψ = 22°-ϑ = 45° and Ψ = 40°-ϑ = 45°, attractive magnetic forces were produced along the direction of the precession axis (0°≤ β ≤30°) and repulsive forces in orthogonal directions (30°≤ β ≤90°) **(Figs. 1d, e)**. Magnetic interaction force decays with distance among microrobots as $\sim r^{-4}$ and its magnitude scales with $\sim B^2 N^2 a^2$, where $N$ is the number of beads in a chain and $a$ is the radius of the magnetic particles **(Fig. 1e, see SI for details)**. These results show that physical interactions among individual microrobots can be engineered by tuning precession and tilt angles of applied magnetic fields.

**Motility and steering of individual microrobots**

Upon application of a tilt angle to the precession axis, rotating microrobotic chains can locomote on the surface and direction of locomotion can be controlled by changing the tilt orientation of the precession axis **(Fig. 2a, Movie S2)**. Velocity of individual chains increases with the length of the chain, actuation frequency, and applied tilt and precession angles **(Fig. 2b)**. Experimental results indicated that change in precession angle had a greater influence on the chain velocity compared to the tilt angle. These results show that locomotion of individual microrobotic chains can be guided and their velocities can be tuned by changing the chain length and frequency.

Propulsion of microrobotic chains is mediated by mismatch of hydrodynamic mobility between two ends of the chain, which is due to the presence of a no-slip surface. Presence of a surface hinders motion of particles closer to the surface, which creates a gradient in hydrodynamic mobility of particles[25]. As a result, in-plane component of torque on a chain induced by magnetic field is translated into linear motion. Indeed, numerical modeling of the chain dynamics, incorporating effects of magnetic interactions, hydrodynamics including wall effects, gravity, and solid body collisions, showed that chain translation is plausible based on the proposed mechanism

**(see SI for details)**. Translation velocity increases almost linearly with angular frequency and number of beads in a chain, which matches with the experimental observations. Furthermore, simulated chain velocities were in agreement with the experimental observations for all tilt and precession angle combinations for different frequencies **(Fig. 2b)**.

**Collective order in microrobotic swarms**

Pairwise interactions among individual microrobots and their motility determine the collective behavior when greater number of microrobots are present in a swarm configuration. We characterized velocity of microrobotic swarms constructed using varying precession and tilt angles **(Fig. 3a, Fig. S2)**. Mean velocities of microrobot swarms matched with individual chain velocities actuated using same precession and tilt angles. However, the distribution of chain velocities among the population was narrow for configurations of $\Psi = 45°$-$\vartheta = 22°$, $\Psi = 30°$-$\vartheta = 30°$, $\Psi = 22°$-$\vartheta = 45°$, whereas a broad range of velocities was observed for configuration of $\Psi = 40°$-$\vartheta = 45°$ **(Fig. 3a, Fig. S2)**.

Distribution of microrobot morphologies, measured in number of beads per microrobot, in a swarm population strongly depends on applied tilt and precession angles **(Figs. 3b, c)**. A narrower distribution of microrobot morphologies was observed in swarms formed at $\Psi = 45°$-$\vartheta = 22°$ and $\Psi = 30°$-$\vartheta = 30°$, with almost all of the chains composed of 3-5 beads. However, swarms formed at $\Psi = 22°$-$\vartheta = 45°$ and $\Psi = 40°$-$\vartheta = 45°$ displayed a broader distribution of microrobot morphologies, displaying a mixture of short and long chains, as well as thicker aggregates **(Fig. 3c, Fig. S3, Movie S3)**. Actuation frequency was observed to have little effect on distribution of chain morphologies **(Fig. 3d, Fig. S3).**

We further characterized spatial distribution of microrobotic swarms using nearest neighbor analysis **(Fig. 3e, Fig. S4)**. Swarms formed at $\Psi = 30°$-$\vartheta = 30°$ showed a narrower distribution of nearest neighbor distances, indicating an even spatial distribution of microrobots, compared to swarms formed at $\Psi = 40°$-$\vartheta = 45°$. This discrepancy in spatial distribution between $\Psi = 30°$-$\vartheta = 30°$ and $\Psi = 40°$-$\vartheta = 45°$ was further visualized using Voronoi diagrams **(Fig. 3f)** and density field plots **(Fig. 3g)**. These results show that microrobots with unidirectional repulsive interactions (*i.e.*, $\Psi = 45°$-$\vartheta = 22°$ and $\Psi = 30°$-$\vartheta = 30°$) form swarms with narrower distributions of morphology, thus resulting in well-controlled population characteristics including velocity and spatial organization over large time scales (> 30 min) and distances (~1 cm). On the other hand, presence of directional attractive interaction among chains in a swarm leads to aggregation of smaller chains into larger aggregates with less-defined morphologies, leading to a larger variance in velocity and spatial distributions.

**Locomotion of microrobotic swarms through confined environments**

Adaptation of microrobotic swarms to their confined local environment, without compromising their integrity and functionality, is a crucial step towards real-world applications. To test dynamic adaptation of our microrobot swarms to the constraints in their environment, they were guided through an array of convex obstacles forming varying confined spaces in between **(Fig. 4, Fig. S5, Movie S4)**. When traversing obstacles, the swarm was compressed in narrowing spaces and displayed expansion once past the obstacles, demonstrating compressibility of swarms **(Fig. 4a)**. Individual microrobots colliding with the obstacle walls were observed to slide on the walls in counter-clockwise direction and, once the contact with the wall was lost, continued freely in the direction of magnetic propulsion **(Fig. 4b)**. Quantification of microrobot density over discrete time intervals showed increased densities at the confined regions, particularly around the periphery of the obstacles **(Fig. 4c)**. An increase in density was also evident in the distribution of nearest neighbor distances in the confined region compared to the control without any geometric confinements **(Fig. 4d)**. Furthermore, comparison of chain area histograms for control and confined regions indicated preserved chain morphologies when the swarm was traversing obstacles **(Fig. 4e)**. These results show that microrobotic swarms with engineered inter-robot physical interactions can navigate through simple porous environments (*e.g.*, arrays of convex obstacles), while preserving their structural and functional integrity. However, the gap distance between obstacles cannot be smaller than the individual microrobot length for robot's structural integrity.

**Controlled cargo transport by mobile microrobotic swarms**

Compared to individual microrobot units, collective microrobotic swarms can accomplish tasks in parallel that can increase their functional throughput. To demonstrate parallel cargo transportation capability of microrobotic swarms, we introduced large (5-20 µm particles) and small (1 µm tracer particles) cargoes among arrays of microrobots **(Fig. 5, Movie S5)**. Calculated trajectories of cargoes **(Fig. 5a)** showed a transport velocity up to 2 µm/s **(Fig. 5b)**, while microrobots maintained an average translational velocity of 3 µm/s and 4 µm/s when actuated at 3 Hz and 5 Hz, respectively **(Fig. 5b)**. In addition, polar distribution histograms for chain and cargo translation direction were overlapping, indicating directional transport of cargoes near the surface **(Fig. 5c)**. When microrobot swarms were mixed with small 1 µm tracer particles, quantified trajectories **(Fig. 5d)** displayed an average translational velocity of 5 µm/s for chains actuated at 5 Hz, while the average transport velocity of tracer particles was around 3 µm/s **(Fig. 5e)**. Similar to cargo transport near the surface, transport of tracer particles in the bulk fluid was also directional, as shown by polar distribution histograms for chain and cargo translation directions **(Fig. 5f)**. These results show that when chain microrobots are actuated in high numbers in a swarm configuration, directional transport of large cargoes near the surface and small cargoes in the bulk fluid can be achieved. Such transport phenomena using mobile chain microrobots are reminiscent of cilia found in living systems, which have also inspired fabrication of biomimetic artificial cilia[26,27]. Parallel transportation and distribution of a large number of cargoes near a wall surface and in bulk fluid using microrobotic swarms can be significant in future applications of drug distribution and coverage in wide body cavities and vasculature.

**Discussions**

In nature, communication and cooperation of individual organisms, such as cells, fishes or insects, in large groups facilitate achievement of complex tasks that would be impossible for the individuals. More interestingly, local communication between limited individuals, which do not possess the higher order information of the group, can give rise to complex global behavior that is vital for navigation, preying, foraging, and survival[28]. Inspired by nature, robotic swarms have been developed at centimeter scales, which allow decentralized simple and modular units to be reconfigured into a team via local interactions achieved by using on-board microcontrollers and infrared sensors[29]. Despite advances in computational techniques, which could be advantageous at the macro scale, miniaturization of such approach down to microscale swarms presents significant challenges, due to lack of on-board microcontrollers, powering, and sensors on current mobile microrobots[10]. Therefore, in design of microrobotic swarms, local communication of individual units need to rely on physical interactions, which is also the case for some biological swarms, such as hydrodynamic interactions among swimming bacteria[30]. Here, we report an example of such a microrobotic swarm with well-defined collective order enabled by physical communication of individual robots via engineered magnetic interactions.

The microrobots described here were constructed by dynamic self-assembly of magnetic particles with applied rotating magnetic fields. Propulsion mechanism of these self-assembled microrobots relied on symmetry breaking by exploiting the nearby surface. Other microrobots utilizing similar mechanisms have been reported previously, including surface walkers[21], microwheels[18,22], kayaks[31], spinning aggregates[32], and nanorod-sphere propellers[23]. Although these microrobots were able show advanced features, including fast propulsion, flow generation, and rolling against gravitational forces, collective operation of such microrobots in high concentrations may not be feasible due to formation of large aggregates, because of unintended magnetic interactions among units. Precise control over magnetic field precession in this study allows engineering magnetic interactions between each unit, which prevents aggregation of microrobots even when compressed as a group in confined spaces.

Engineered interactions among microrobots described here mainly rely on magnetic repulsion of microrobots, which we showed to be necessary for maintaining the microrobot integrity. Purely repulsive interactions would eventually lead to spreading of robots over long time scales in open spaces. However, such spreading may be especially useful when operating in confined environments to spread and cover large surface areas, such as in bodily cavities. Furthermore, to enhance cohesion of swarms, attractive interactions can be further introduced into the design by using additional decoupled external fields, such as acoustic and electrical fields. By fine selection and tuning of pairwise interactions based on different physical effects, attractive and repulsive forces can be balanced at specific distances, which would result in tightly defined spatial ordering of microrobot clusters[33-35].

Microrobotic swarms with a collective order can enable propulsion over large distances while preserving structure and functions of individual units. Moreover, dynamic adaptation of microrobotic swarms to confined spaces is crucial for navigation in unstructured environments, such as blood vessels. Therefore, microrobotic swarms engineered using the design approach described here can hold potential for real-world medical applications, including active transport and diffusion of drug molecules and remote heating-based microsurgery. Overall, we describe a design methodology for designing microrobot swarms with collective order by exploiting physical interactions among individual units.

**Materials and Methods**

*Experimental Setup*

To characterize dynamic self-assembly and motility of chain arrays, a custom five-coil magnetic guidance system placed on an inverted optical microscope (Zeiss Axio Observer A1, Carl Zeiss, Oberkochen, Germany). The coil setup was designed to generate 20 mT in *x*- and *y*-directions, as well as 10 mT in *z*-direction (out of plane)[2]. Individual coils were controlled independently via a current controller (Escon 70/10, Maxon Motor AG) and current values were determined by the pre-calibrated field to current ratios. The generated magnetic field was measured to be uniform within 5 mm from the center of the workspace. In addition, magnetic guidance system housed a microfluidic channel (75 μm height x 6 mm width x 10 mm length) composed of laser cut poly(methyl methacrylate) (PMMA) pieces, encompassing an inlet and an outlet, and double sided tape, defining channel shape and height, attached to a cover glass[36]. To test locomotion of microrobot swarms through confined spaces, obstacles with a diameter of 120 μm diameter were patterned on cover glasses using a commercial two-photon lithography platform (Nanoscribe, Eggenstein-Leopolds-hafen, Germany).

*Self-assembly of microrobot swarms*

Superparamagnetic polystyrene microparticles with around 5 μm diameter (Sigma Aldrich, St Louis, MO) were utilized for self-assembly of chain arrays. The particles were suspended in 0.1% Tween 20 solution (Sigma Aldrich, St Louis, MO) to prevent any non-specific aggregations and injected into microchannels. Then, an out of plane magnetic field (in *z*-axis) was applied to uniformly disperse the microparticles followed by an in plane rotating magnetic field resulting in self-assembled chains on the surface. Then, pre-determined tilt and precession angles were applied to the rotating magnetic field, producing mobile microrobot chain arrays with specific inter-robot attractive or repulsive forces.

*Data Analysis*

Acquired images were processed using Fiji[37] to identify individual chains, positions, and projected surface areas. A tracking software[38] was used to reconstruct trajectories of individual chains and their velocities. Nearest neighbor, based on Delaunay triangulation, and field density analyses were performed using custom-written scripts in MATLAB (MathWorks, Natick, MA). Density field $\rho$ is obtained by averaging binarized pixel intensities over a window with a diameter of two mean nearest neighbor distances, and is normalized as $\rho^* = \rho/\bar{\rho} - 1$.

*Modeling pairwise interactions and chain propulsion*

For calculation of magnetic interaction forces among microrobot units, we considered two self-assembled chains, each consisted of *N* paramagnetic beads with radius *a* and magnetic susceptibility $\chi$. Under an applied field *B*, a magnetic dipole moment $\boldsymbol{m}$ is induced for each bead. Two chains interact with each other via magnetic forces $\boldsymbol{F}$. Pairwise attractive and repulsive interaction forces $F_p$ are quantified by taking the component of $\boldsymbol{F}$ acting along the line crossing centers of chains. Two chains separated by one chain length distance between their centers interact with a characteristic magnetic force $F_0 = \frac{\pi}{12\mu_0}\left(\frac{a\chi B}{N}\right)^2$, which is used for normalizing $\boldsymbol{F}$ and $F_p$ in Figs. 1d-e **(see SI for calculation of $F_0$)**.

Chain propulsion velocities are simulated by modeling the dynamics of individual superparamagnetic beads under external fields via

$$\dot{r}_i = M_{ij} \cdot (F_j^M + F_j^B + F_j^W + F_j^G)$$

where beads interact with magnetic dipole forces ($F^M$), particle-particle ($F^B$) and particle-surface ($F^W$) excluded volume forces, and gravitational ($F^G$) forces. The grand mobility tensor $M$ couples the velocities of beads ($\dot{r}_i$) to the forces acting on each bead through contributions of self and pair hydrodynamic mobility tensors including hydrodynamic interactions with the wall surface[39].

**Acknowledgements**

The authors would like to thank J. Giltinan for his assistance in magnetic coil setup and V. Kishore for his helpful discussions. Y.A. thanks Alexander von Humboldt Foundation for the Humboldt Postdoctoral Research Fellowship. This work is funded by the Max Planck Society.

**Competing financial interests**

The authors declare no competing financial interests.

13. Yim S, Sitti M. SoftCubes: Stretchable and self-assembling three-dimensional soft modular matter. *The International Journal of Robotics Research* **33**, 1083-1097 (2014).

14. Carlsen RW, Sitti M. Bio-hybrid cell-based actuators for microsystems. *Small* **10**, 3831-3851 (2014).

15. Ricotti L, *et al.* Biohybrid actuators for robotics: A review of devices actuated by living cells. *Science Robotics* **2**, (2017).

16. Martel S, Mohammadi M. Using a swarm of self-propelled natural microrobots in the form of flagellated bacteria to perform complex micro-assembly tasks. In: *2010 IEEE International Conference on Robotics and Automation* (ed^(eds) (2010).

17. Yan X, *et al.* Multifunctional biohybrid magnetite microrobots for imaging-guided therapy. *Science Robotics* **2**, (2017).

18. Tasci TO, *et al.* Enhanced Fibrinolysis with Magnetically Powered Colloidal Microwheels. *Small* **13**, (2017).

19. Wang B, *et al.* Reconfigurable Swarms of Ferromagnetic Colloids for Enhanced Local Hyperthermia. *Advanced Functional Materials* **0**, 1705701 (2018).

20. Zhuang J, Wright Carlsen R, Sitti M. pH-Taxis of Biohybrid Microsystems. *Scientific Reports* **5**, 11403 (2015).

21. Sing CE, Schmid L, Schneider MF, Franke T, Alexander-Katz A. Controlled surface-induced flows from the motion of self-assembled colloidal walkers. *Proc Natl Acad Sci U S A* **107**, 535-540 (2010).

22. Tasci TO, Herson PS, Neeves KB, Marr DW. Surface-enabled propulsion and control of colloidal microwheels. *Nat Commun* **7**, 10225 (2016).

23. García-Torres J, Calero C, Sagués F, Pagonabarraga I, Tierno P. Magnetically tunable bidirectional locomotion of a self-assembled nanorod-sphere propeller. *Nature Communications* **9**, 1663 (2018).

24. Martinez-Pedrero F, Ortiz-Ambriz A, Pagonabarraga I, Tierno P. Colloidal Microworms Propelling via a Cooperative Hydrodynamic Conveyor Belt. *Physical Review Letters* **115**, 138301 (2015).
16

# Figures

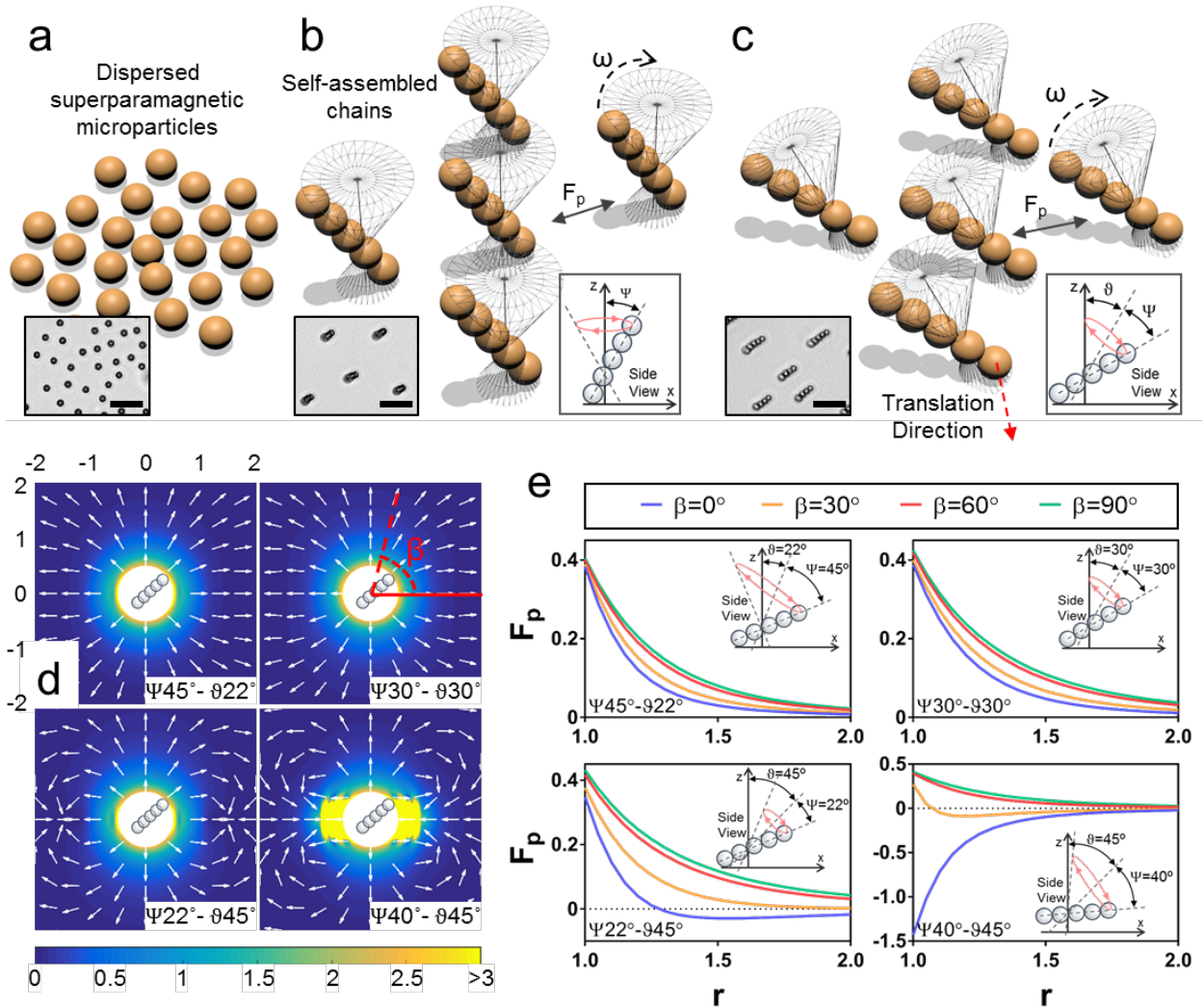

**Figure 1. Design of collective microrobotic swarms. (a)** Magnetic microrobots are composed of superparamagnetic microparticles and **(b)** self-assemble into chains simultaneously in large numbers via an applied rotating magnetic field with a precession angle (Ψ). *In-situ* formed chains rotate around their precession axes with a rotational velocity (ω) and apply magnetically repulsive or attractive forces to one another ($F_p$). **(c)** Self-assembled chains locomote on a solid surface, with a tilt angle (ϑ) applied to the precession axis in the presence of magnetic inter-chain forces. Insets at the bottom-left show typical experimental images illustrated in **(a-c)**. Scale bars are 25 μm. Insets

at the bottom-right demonstrate applied tilt and precession angles from side view. **(d)** Numerical analysis of magnetic dipolar interaction forces between two chains depending on inter-robot positions, averaged over a single rotational cycle. Vectors, indicated by white arrows, shows the direction of the force imposed by the first chain that is located at the origin on a virtual second chain positioned at the vector locations. Color bar shows the magnitude of inter-chain forces, normalized by characteristic chain interaction force $F_0$. Chains depicted in the center of each plot represents the approximate area swept by a single chain. **(e)** Quantification of magnetic forces ($F_p$) acting along the axis between two microrobotic chains, as shown in **(d)**, depending on their distance (r), normalized by one chain length, and angular direction ($\beta$). Positive $F_p$ values indicate repulsion, whereas negative ones indicate attraction between units. Insets in the plots show prescribed tilt and precession angles for each analysis.

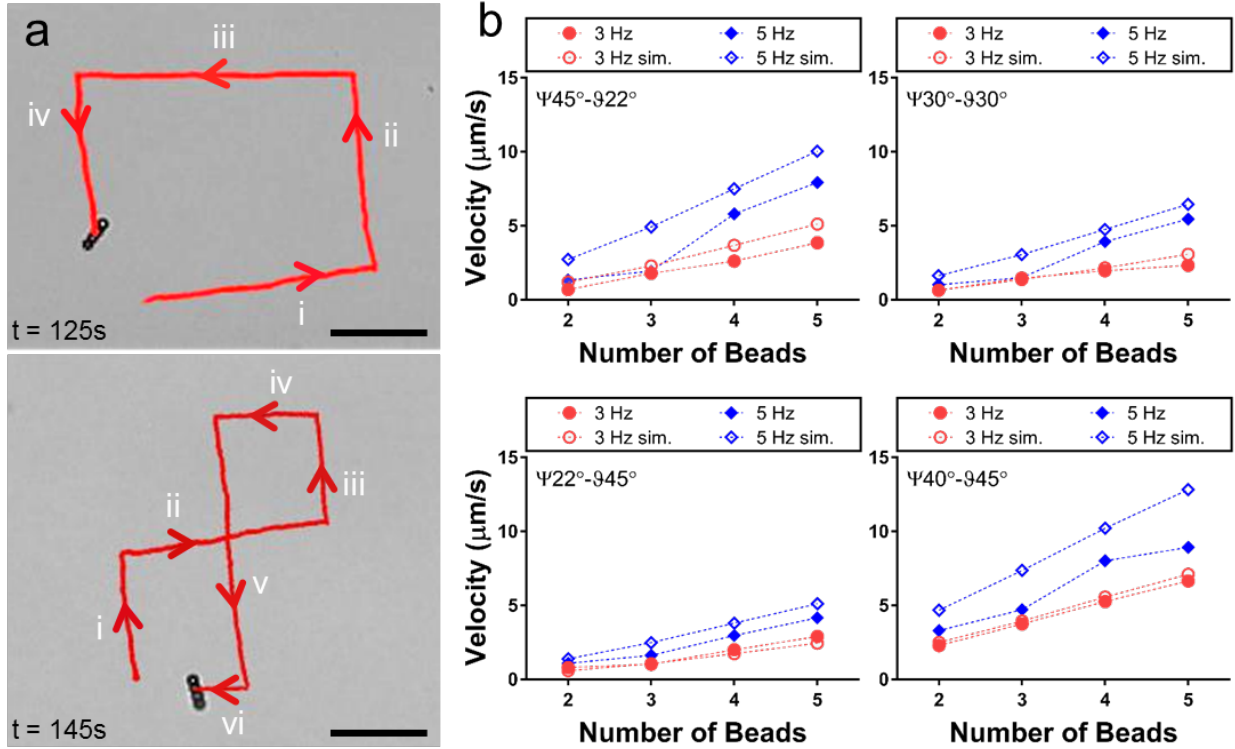

**Figure 2. Characterization of single microrobot motility. (a)** Propulsion of a single microrobot, consisting of self-assembled microparticles, can be guided by changing orientation of the precession axis. Scale bars are 200 μm. **(b)** Velocity of a single microrobot depends on number of beads in the chain and frequency of the rotating magnetic field, as well as tilt and precession angles. Closed and open symbols correspond to experimental measurements and simulation results, respectively.

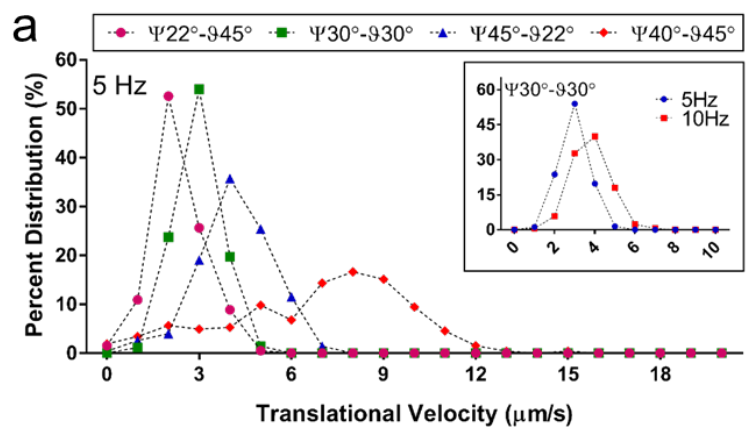
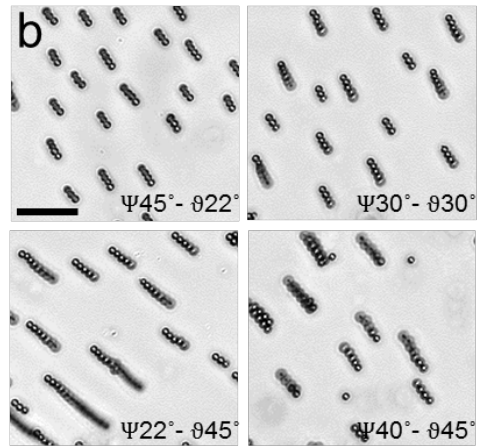
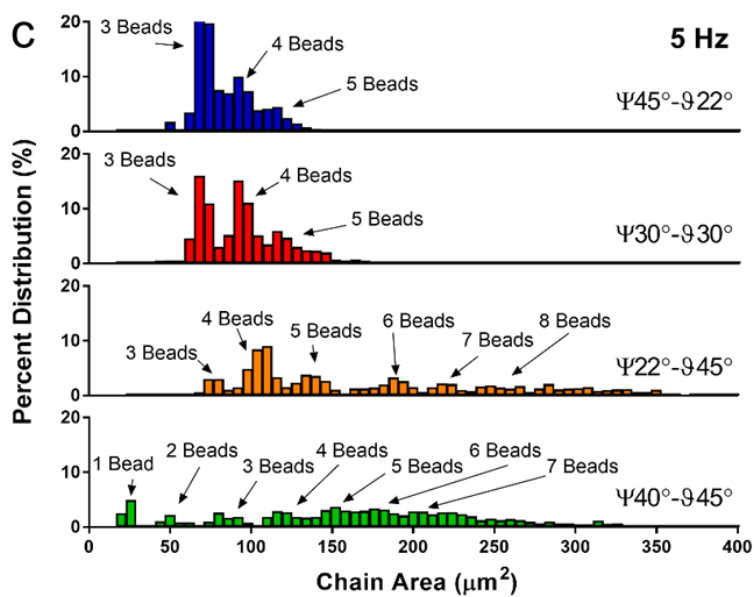
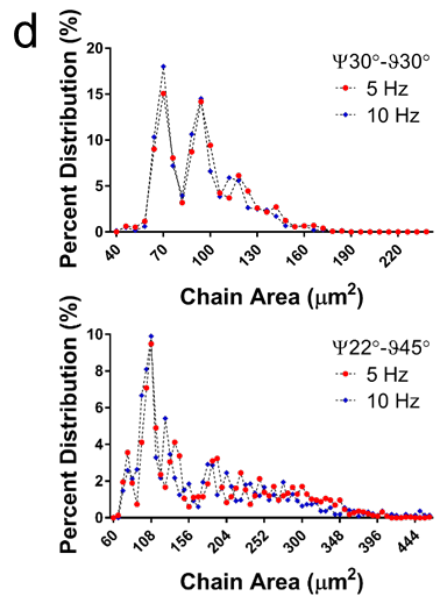
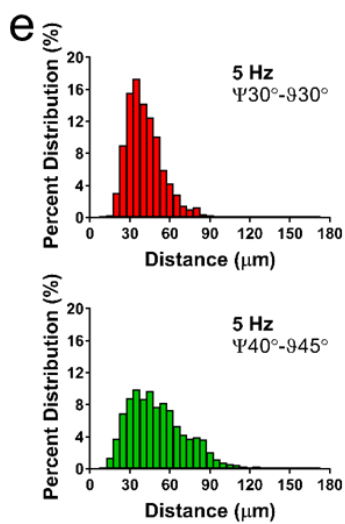
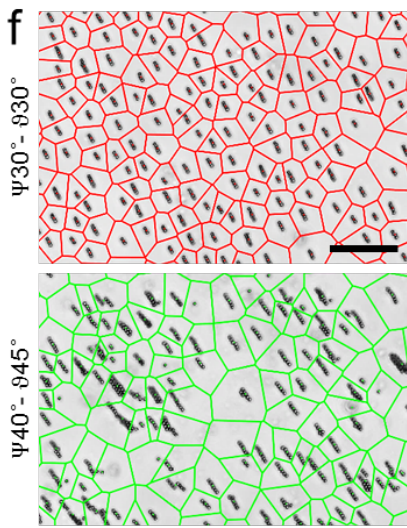
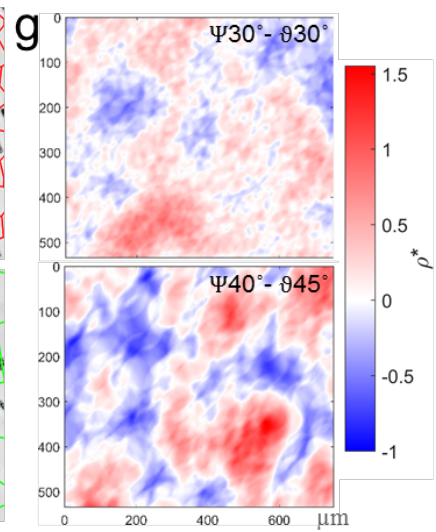

**Figure 3. Collective order in self-assembled mobile microrobotic swarms.** **(a)** Velocity histogram of microrobotic swarms at varying tilt and precession angles. Inset shows shift in velocity histogram with increased frequency. **(b)** Typical microscopy images of microrobotic swarms formed at varying tilt and precession angles. Scale bar is 50 μm. **(c)** Percent distributions of chain area in microrobotic swarms formed at different tilt and precession angles. Chain area corresponds to number of beads in a microrobotic unit, which is indicated over histogram peaks in the plots. Microrobotic swarms formed at $\Psi = 45°$-$\vartheta = 22°$ and $\Psi = 30°$-$\vartheta = 30°$ display a narrower distribution of microrobot morphologies compared to swarms formed at $\Psi = 22°$-$\vartheta = 45°$ and $\Psi = 40°$-$\vartheta = 45°$. **(d)** Comparison of chain area histograms for microrobotic swarms actuated at 5 Hz and 10 Hz. **(e)** Nearest neighbor distribution, as a measure of inter-robot separation, showed a more distinct and narrow peak for microrobotic swarms formed at $\Psi = 30°$-$\vartheta = 30°$ compared to $\Psi = 40°$-$\vartheta = 45°$, with a broader spectrum. **(f)** Voronoi diagrams and **(g)** normalized density field ($\rho^*$) showed a more uniform spatial distribution for microrobotic swarms formed at $\Psi = 30°$-$\vartheta = 30°$ compared to $\Psi = 40°$-$\vartheta = 45°$. Zero corresponds to the mean. Scale bar is 50 μm.

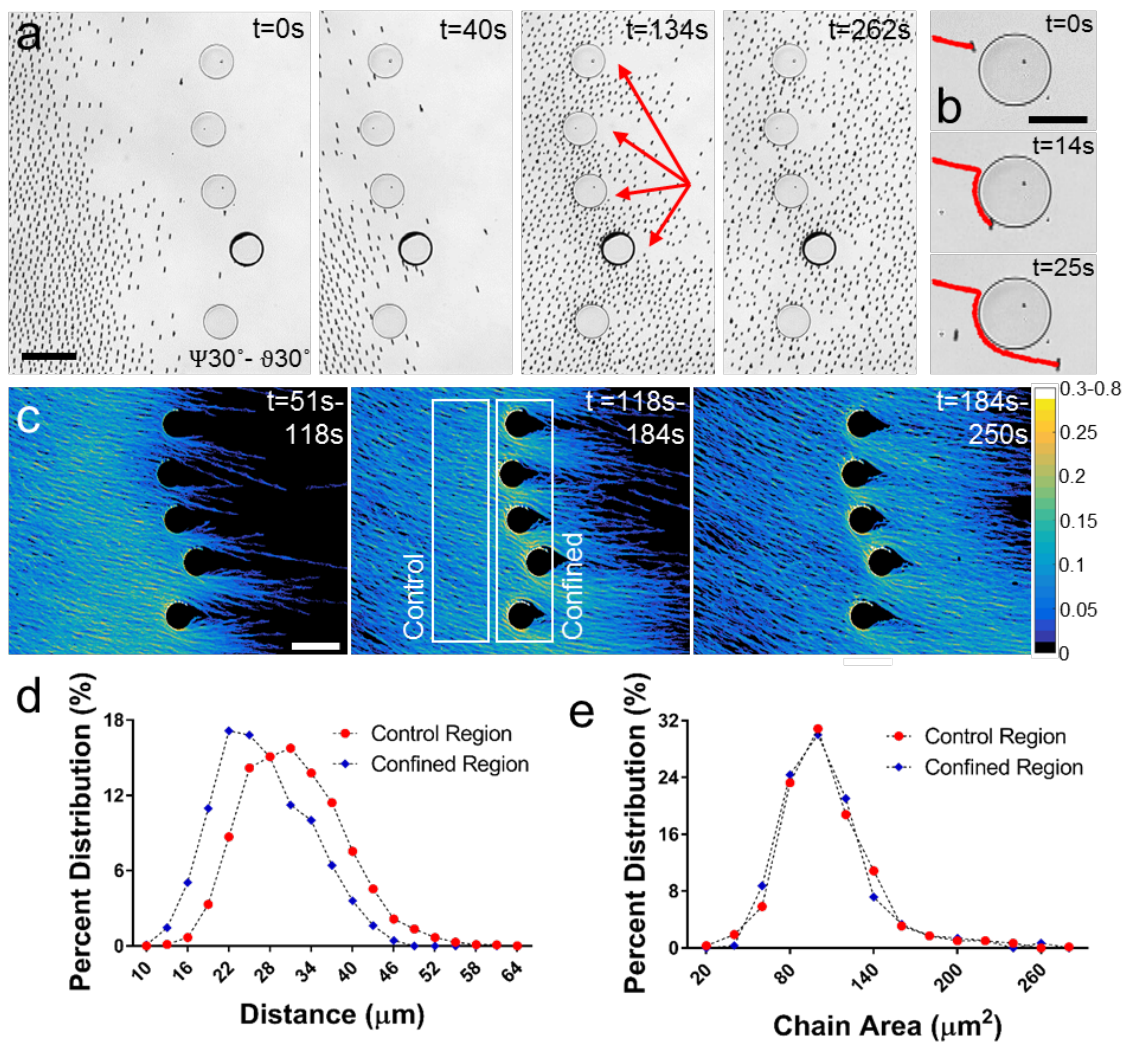

**Figure 4. Locomotion of microrobotic swarms through confined environments. (a)** Microrobotic swarm traversing an array of obstacles displays a compressible behavior, where the swarm is compressed in confinement and expands after obstacles. Microrobots form wakes behind obstacles indicated with red arrows. Scale bar is 200 μm. **(b)** Individual chains, when encountered with the wall boundary, slip over the obstacle in counter-clockwise direction and continue in the direction of magnetic actuation when the contact is lost. Scale bar is 100 μm. **(c)** Measurement of microrobot density averaged over discrete time intervals reveal increased swarm density in narrowing confinements. Scale bar is 250 μm. **(d)** Nearest neighbor analysis indicates increased

microrobot density in confined region compared to control. **(e)** Comparison of chain area histograms for control and confined regions reveals that chain morphologies are preserved when the swarm is travelling through obstacles.

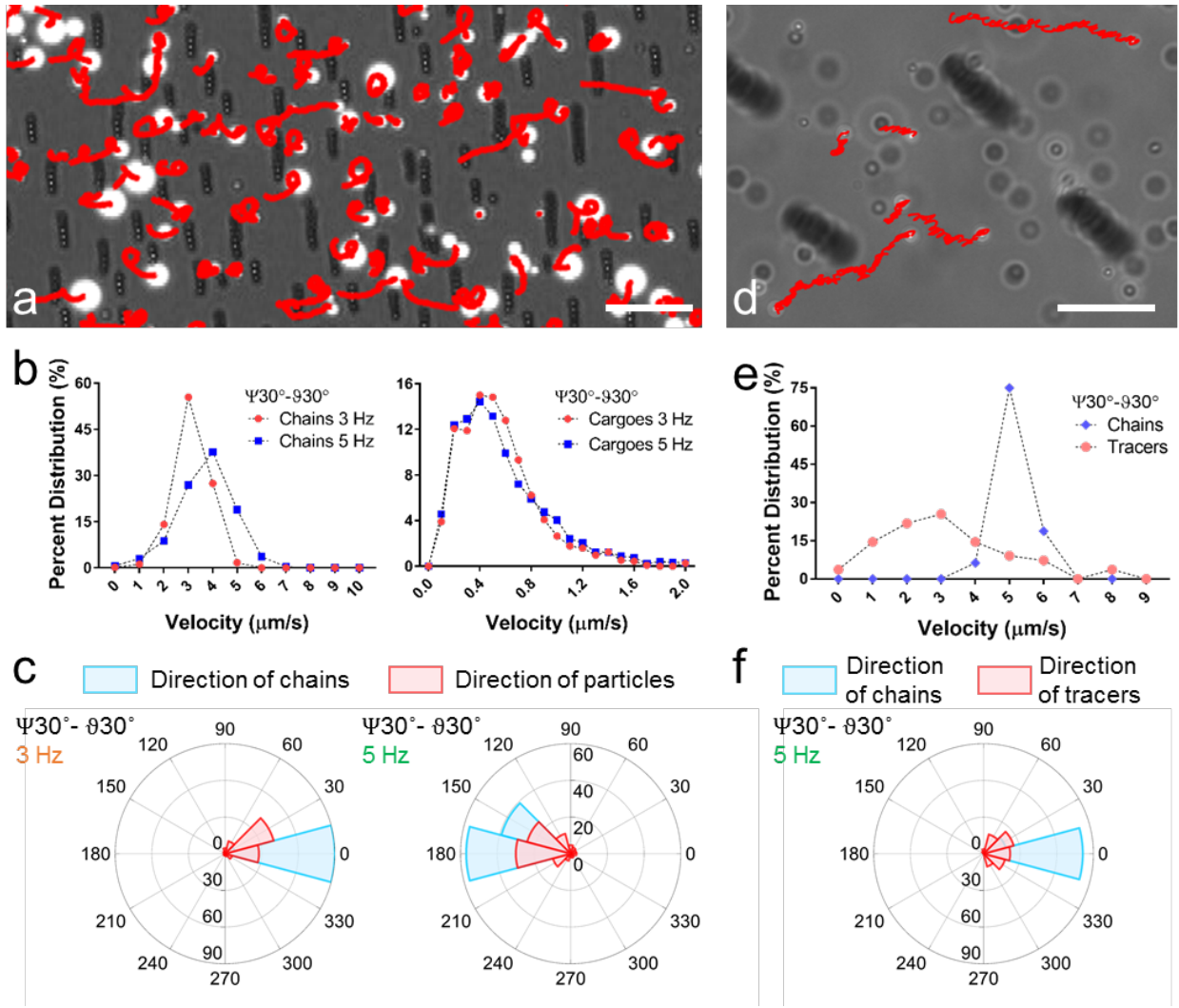

**Figure 5. Surface and bulk fluid transport induced by mobile microrobotic swarms. (a)** Trajectories of large cargoes (5-20 μm particles) transported by the microrobot swarm. Scale bar is 50 μm. **(b)** Velocity histograms for both microrobotic chains and cargoes when chains were actuated at 3 Hz and 5 Hz. **(c)** Polar histograms for microrobotic chains and cargoes demonstrate directed transport of cargoes in the same direction of chain locomotion at 3 Hz and 5 Hz. **(d)** Non-contact transport of small cargoes (1 μm tracer particles) in bulk fluid by flows generated by surface propulsion of microrobotic swarm. Scale bar is 20 μm. **(e)** Velocity histograms for microrobot swarm and tracer particles. **(f)** Polar histogram for microrobot chains and tracer particles shows directed transport of cargoes in bulk fluid.

26